\documentclass[a4paper,twoside]{article}

\usepackage{epsfig}
\usepackage{subcaption}
\usepackage{calc}
\usepackage{amssymb}
\usepackage{amstext}
\usepackage{amsmath}
\usepackage{amsthm}
\usepackage{multicol}
\usepackage{pslatex}
\usepackage{apalike}
\usepackage{algorithm2e}
\usepackage[bottom]{footmisc}
\usepackage{verbatim}
\usepackage{hyperref}
\usepackage{booktabs}
\usepackage{array}
\usepackage{float}
\usepackage{stfloats}
\usepackage{makecell}
\usepackage{threeparttable}
\usepackage{footnote}

\usepackage{SCITEPRESS}  

\newcolumntype{C}[1]{>{\centering\arraybackslash}p{#1}}

\begin{document}

\title{MOSAIC-FL, a micro-service based privacy-preserving framework with application to genomics}

\author{\authorname{
Paul Largillier\sup{1}\orcidAuthor{0009-0009-0218-3457},
Karl Paygambar\sup{2}\orcidAuthor{0009-0001-2768-1569},
Cédric Gouy-Pailler\sup{1}\orcidAuthor{0000-0003-1298-7845},
Vincent Meyer\sup{2}\orcidAuthor{0000-0001-9144-1618},
Mallek Mziou\sup{2}\orcidAuthor{0000-0001-8324-0891}
and Oana Stan\sup{1}\orcidAuthor{0000-0003-2419-9037}
}
\affiliation{\sup{1}Université Paris-Saclay, CEA, LIST, France}
\affiliation{\sup{2}Université Paris-Saclay, CEA, CNRGH, France}
\email{
\{paul.largillier, karl.paygambar, cedric.gouy-pailler,
vincent.meyer, mallek.mziou, oana.stan\}@cea.fr
}
}


\abstract{Security and privacy are primordial requirements for Federated Learning (FL), especially in fields such as healthcare and genomics when sensitive information has to be analyzed. Our FL framework is designed to address these challenges while proposing a modular, flexible and micro-service architecture. More precisely, it integrates an efficient gRPC communication layer and a Finite State Machine to ensure robust component synchronization and threat detection, and it is based on a fault-tolerant secure aggregation protocol using a Threshold variant of the CKKS homomorphic cryptosystem. This allows blind model aggregation by an orchestration server, requiring a minimum of $t$-out-of-$N$ active clients for decryption while minimizing communication overhead thanks to both cryptographic and network protocols. We demonstrate the framework's effectiveness through diverse use cases, ranging from standard image recognition (EMNIST) to complex genomic classification including breast cancer subtyping on TCGA, evaluating system performance across different threshold values and model scales.}

\onecolumn \maketitle \normalsize \setcounter{footnote}{0} \vfill

\section{\uppercase{Introduction}}
\label{sec:introduction}

Federated Learning (FL) \cite{McMahan2017} is a collaborative approach enabling multiple entities to train a common Machine Learning (ML) model in a distributed way without sharing their sensitive data.
In its centralized form, it relies on a global server that updates the global model by federating the client's model updates using different aggregation rules (the most common one being the federated averaging).
The original FL approach has several drawbacks in terms of security and privacy. Various research results showed numerous vulnerabilities (e.g. model inversion attacks, membership inference attacks, property inference attacks).
In the literature, multiple privacy-preserving techniques have been proposed to mitigate these risks. However, deploying secure solutions can be constrained by the architectures of existing systems. In addition, many research-oriented frameworks are highly specialized, making them difficult to scale and adapt to real-world deployments. We argue that a practical FL solution must jointly address security, model diversity, deployment heterogeneity, and robust monitoring, requirements that existing frameworks only partially fulfill. As such, in this paper we propose MOSAIC-FL (Modular and Secure Federated Learning via Microservice Orchestration) a privacy-preserving federated framework that features optimized communication protocols, a formally specified architecture based on state machines, a fault-tolerant secure protocol using Threshold Fully Homomorphic Encryption (ThFHE), and a language-agnostic design that facilitates the integration of new security functionalities and machine-learning models.

The main contributions of this work are summarized as follows:
\begin{itemize}
    \item A modular and extensible micro-service-based architecture with efficient gRPC communication protocols allowing easy integration of different ML frameworks and cryptographic libraries;
    \item A Finite State Machine (FSM) that synchronizes the main software components and eventually detects and prevents failures and malicious behaviors;
    \item A fault-tolerant protocol for secure aggregation based on the CKKS implementation of the ThHE scheme of Mouchet et al.~\cite{mouchet2023}, allowing the orchestration server to blindly compute the aggregation of the models and to decrypt the results only if $t$ out of the $N$ clients participating in the training are online and active.
    \item Experimental validation of the framework through several use cases, including standard image workloads (EMNIST) and realistic, more complex genomic classification tasks. System-level performance, in terms of communication overhead and training accuracy, is evaluated across various federated configurations. 
\end{itemize}

\section{\uppercase{Related Works}}
\label{sec:relatedWork}
As illustrated in Table~\ref{tab:fl_comparison}, which compares our framework with the most popular FL frameworks, there is a clear distinction in how security and architecture are handled across the state of the art. While established frameworks like Flower and PySyft provide robust infrastructures, they typically lack native support for threshold homomorphic encryption. Most existing open-source platforms rely on single-key variants of homomorphic cryptosystems that require a trusted aggregator or a central authority to hold the decryption key. This reliance introduces a single point of failure and limits the applicability of these frameworks in contexts where participants cannot be fully trusted. The architectural design of mainstream frameworks also presents limitations regarding modularity and language interoperability. FATE, for instance, adopts a monolithic design that can be rigid and difficult for researchers to extend in order to test new security primitives. While PySyft offers modularity through its domain-based design, it remains deeply rooted in the Python ecosystem, making it challenging to integrate with low-level languages like C++ for high-performance cryptography tasks. Flower addresses some of these issues with a more agnostic approach but still requires the development of specific software development kits to support different platforms and languages. The management of client dropouts remains a persistent challenge in current implementations. Most frameworks employ fail-stop or simple retry-based mechanisms that struggle with the complex synchronization required for cryptographic protocols. Specifically, the lack of native ThHE support in these frameworks means they cannot perform collaborative decryption if a subset of clients becomes unavailable during a round. Finally, the polyglot nature achieved through the use of gRPC removes the need for a dedicated SDK, representing an advantage over frameworks primarily limited to a single ecosystem.

\begin{table*}[t]
\caption{Comparison of MOSAIC-FL with popular FL frameworks.
  $^1$\url{https://www.tensorflow.org},
  $^2$\url{https://pytorch.org},
  $^3$Lattigo: \url{https://github.com/tuneinsight/lattigo}}
\label{tab:fl_comparison}
\footnotesize
\renewcommand{\arraystretch}{1.3}
\begin{tabular}{|l|C{2.8cm}|C{2.6cm}|C{2.6cm}|C{2.6cm}|}
\hline
\textbf{Criteria}
  & \textbf{MOSAIC-FL}
  & \href{https://github.com/adap/flower}{\textbf{Flower}}
  & \href{https://github.com/FederatedAI/FATE}{\textbf{FATE}}
  & \href{https://github.com/OpenMined/PySyft}{\textbf{PySyft}} \\
\hline
ML Frameworks
  & \makecell{TensorFlow$^1$ \\ PyTorch$^2$}
  & Agnostic
  & \makecell{PyTorch$^2$ / TF$^1$ \\ Scikit-Learn}
  & \makecell{PyTorch$^2$ \\ TensorFlow$^1$} \\
\hline
Communication Strategy
  & Sync / Async
  & \makecell{Sync (default) \\ Async}
  & Sync
  & Sync / Async \\
\hline
Single-key HE
  & CKKS/BGV/BFV$^3$
  & TenSEAL plugin
  & Paillier \cite{paillier}
  & TenSEAL \cite{tenseal2021} \\
\hline
Threshold HE
  & CKKS/BGV/BFV$^3$
  & None
  & None
  & None \\
\hline
Extensibility
  & \makecell{Very High \\ (Microservices)}
  & \makecell{High \\ (Modular)}
  & \makecell{Medium \\ (Monolithic)}
  & \makecell{High \\ (Domain-based)} \\
\hline
Communication
  & gRPC / REST
  & gRPC / REST
  & gRPC / HTTP
  & gRPC / HTTP \\
\hline
Languages
  & \makecell{Polyglot \\ (Micro-services)}
  & \makecell{Python, Swift \\ Kotlin, TS}
  & \makecell{Python \\ Java, C++}
  & Python \\
\hline
\end{tabular}
\end{table*}

\section{\uppercase{Preliminaries}}
\label{sec:preliminaries}
\subsection{Federated Learning setup}
Federated Learning (FL)
is a decentralized framework that enables multiple \emph{clients} to collaboratively train a shared global model under the orchestration of a central server while keeping the training data localized on the client devices. This helps to protect data privacy while reducing communication costs. After a common  arbitrary initialization of the global model, the FL process consists of successive rounds of communication between the server and the clients.
The most common approach to optimization for FL is the Federated Averaging algorithm \cite{McMahan2017}. At the beginning of each round, the server selects a subset of clients to take part in training for this round; we call these particular clients the \emph{participants}. The server sends the current global model to the participants and each of them trains the model locally with several epochs of stochastic gradient descent (SGD) using its own data. The participants then communicate only the updated parameters or the updates\footnote{Difference between the updated parameters and the old ones.} themselves (depending on the setting) back to the server. Finally, the server computes the weighted average of these updates before accumulating them into the global model, thereby concluding the round.

\subsection{Threshold Homomorphic Encryption}
\label{subsec:the}
Several cryptographic schemes investigate using
multi-key techniques to adapt HE to multi-user settings \cite{survey_mkhe_aloufi},
such as Threshold Homomorphic Encryption (ThHE) \cite{thhe_first,thhe_access_structure_lwe,thhe_universal_thresholdizer,thhe_efficient}, multi-key homomorphic encryption (MKHE) e.g. \cite{mkhe_indep_depth,mkhe_indep_number_users} or more recently, hybrid approaches proposed by \cite{hybrid}.
ThHE schemes allow $N$ entities to
generate a joint global public key and perform a homomorphic computation of an arbitrary function, while allowing a fixed subset $t \leq N$ of these entities to collaboratively decrypt the ciphertexts without revealing their individual partial secret keys.

We decided to use the Threshold multiparty homomorphic encryption scheme of \cite{mouchet2023}, an RLWE (Ring Learning With Errors)-based construction that enables secure evaluation of functions over homomorphically encrypted data while distributing the secret key among $N$ parties $\{P_i\}_{i=1}^N$ with a $t$-out-of-$N$ access structure.
The scheme extends a previous $N$-out-of-$N$ design \cite{Mouchet2020} by introducing a re-sharing procedure based on Shamir secret sharing performed directly over the ciphertext ring.
Among the advantages of this particular synchronous ThHE scheme is that it requires a single round of communication during the setup phase and that the public key can be reused for an unlimited number of aggregations. We instantiated this generic construction for the multiparty version of the CKKS \cite{ckks} scheme.

Let $R_q = \mathbb{Z}_q[X]/(X^n+1)$
and $\chi$ denote an error distribution over $R_q$.
The scheme is defined by the following tuple of algorithms:
\begin{itemize}
    \item $\mathsf{Setup}(1^\lambda, N, t)$.
On input a security parameter $\lambda$ and threshold parameters $(N,t)$,
output public parameters
$
\mathsf{pp} = (R_q, \chi, \Delta, \{\alpha_j\}_{j=1}^N)
$
where $\Delta$ is the scaling factor for CKKS  and  $\{\alpha_j\}$ are distinct public
evaluation points used for Shamir secret sharing over $R_q$.
    \item $\mathsf{DistKeyGen}(\mathsf{pp})$. 
    Each party $P_i$ performs the following steps. First, it samples $s_i \leftarrow R_q$ and error $e_i \leftarrow \chi$. Afterwards, using common randomness, sample $a \leftarrow R_q$ and compute $
    p_{0,i} = -a s_i + e_i \in R_q.
    $
The collective public key is
$
\mathsf{pk} = (p_0, p_1) = \left(\sum_{i=1}^N p_{0,i}, \; a \right)
$
and the collective secret key is additively shared as
$
s = \sum_{i=1}^N s_i.
$
Each party keeps its share $s_i$ private.

    \item $\mathsf{Enc}(\mathsf{pk}, m)$.
To encrypt a plaintext polynomial $m$ (an encoding of the complex data to be encrypted using the scaling factor $\Delta$), sample $u,e_1,e_2 \leftarrow \chi$ and compute
ciphertext $\mathbf{c}=(c_0,c_1)\in R_q^2$ with $c_0 = p_0 u + e_1 + m$ and $c_1 = p_1 u + e_2$.
    \item $\mathsf{Eval}(f,\mathbf{c}_1,\ldots,\mathbf{c}_\ell)$.
Given ciphertexts $c_i$ with $i \in \{1, \ldots, \ell\}$ and a circuit $f$, homomorphic evaluation is performed
using the operations of the RLWE-based HE scheme (here CKKS), producing
$\mathbf{c}_f=f(c_1,\ldots,c_\ell)$.

 \item $\mathsf{ShareReshare}(\{s_i\})$. To obtain a $t$-out-of-$N$ threshold structure, each party $P_i$ constructs
a Shamir sharing of $s_i$: it samples a polynomial $S_i(X)\in R_q[X]$ of degree $t-1$ such that  $S_i(0) = s_i$, for 
  each $j\in \{1,\ldots,N\}$, computes
   $
    s_{i \rightarrow j} = S_i(\alpha_j),
    $
    and sends it privately to $P_j$.
Each party $P_j$ obtains a secondary share
$
\tilde{s}_j = \sum_{i=1}^N s_{i \rightarrow j}.
$
The vector $(\tilde{s}_1,\ldots,\tilde{s}_N)$ forms a Shamir sharing of $s$
with threshold $t$.
\item $\mathsf{PartDec}(\mathbf{c}, \tilde{s}_j)$. Given ciphertext $\mathbf{c}=(c_0,c_1)$, party $P_j$ computes a partial
decryption share
$d_j = c_1 \tilde{s}_j + e'_j,$
where $e'_j \leftarrow \chi$ is fresh noise, and outputs $d_j$.

\item $\mathsf{CombDec}(\{d_j\}_{j \in T}, c_0)$. For any subset $T \subseteq \{1,\ldots,N\}$ with $|T|\ge t$, compute
Lagrange coefficients $\{\lambda_j\}_{j \in T}$ at $0$, recover the partial decryption $d = \sum_{j\in T} \lambda_j d_j \approx c_1 s$ and the plaintext message $m = c_0 + d$. To recover the clear result, one has to decode $m$ (using $\Delta$).
\end{itemize}

\section{\uppercase{Approach}}
\label{sec:approach}
\subsection{Security model}

In the first version of our FL framework we consider that the aggregation server is an honest-but-curious (or semi-adaptive) adversary and that at least $t$ out of the $N$ clients participating in a round of training are also honest-but-curious. In order to address the confidentiality threats coming from the honest-but-curious central server, our framework performs secure aggregation with homomorphic encryption, which enables the update of the global model to be computed in the encrypted domain. After the encrypted local model updates are received from all the $N$ parties, the global model can be decrypted as long as at least $t$ honest clients are online and active.
Moreover, our security model is valid under the hypothesis of non-collusion between the participants in the FL training. Additionally, a requirement of the ThHE scheme we use is the existence of public authentication channels to set up the protocol.

Let us also specify that in the current setting we assume, for simplicity's sake, that in order to perform the homomorphic key generation, the clients had initially performed some kind of consensus protocol to generate the required randomness \cite{Xiao_2020}.

As for the formal security of the underlying threshold homomorphic scheme, we ensure IND-CPA-D security and mitigate recent attacks such as the one from \cite{CPAD-crypto} by proper noise flooding. As specified in \cite{cryptoeprint:2026/031}, synchronous ThHE schemes are additionally vulnerable to a key-recovery attack in which an adversary collects partial decryption shares across different decryptor sets to cancel smudging noise and recover secret key shares via the Chinese Remainder Theorem. To counter this threat, MOSAIC-FL renews the collective key material at every round by re-executing $\mathsf{Setup}$, $\mathsf{DistKeyGen}$, and $\mathsf{ShareReshare}$, which introduces negligible overhead relative to the dominant cost of local training and network communication in the federated learning cycle.

\subsection{Description of the new framework}

The proposed framework departs from the monolithic design patterns prevalent in existing FL libraries. By adopting a microservices architecture, we enforce a strict separation between communication logic, ML computations, and cryptographic primitives.

\begin{figure}[t]  
  \centering
  \includegraphics [width=1\linewidth]{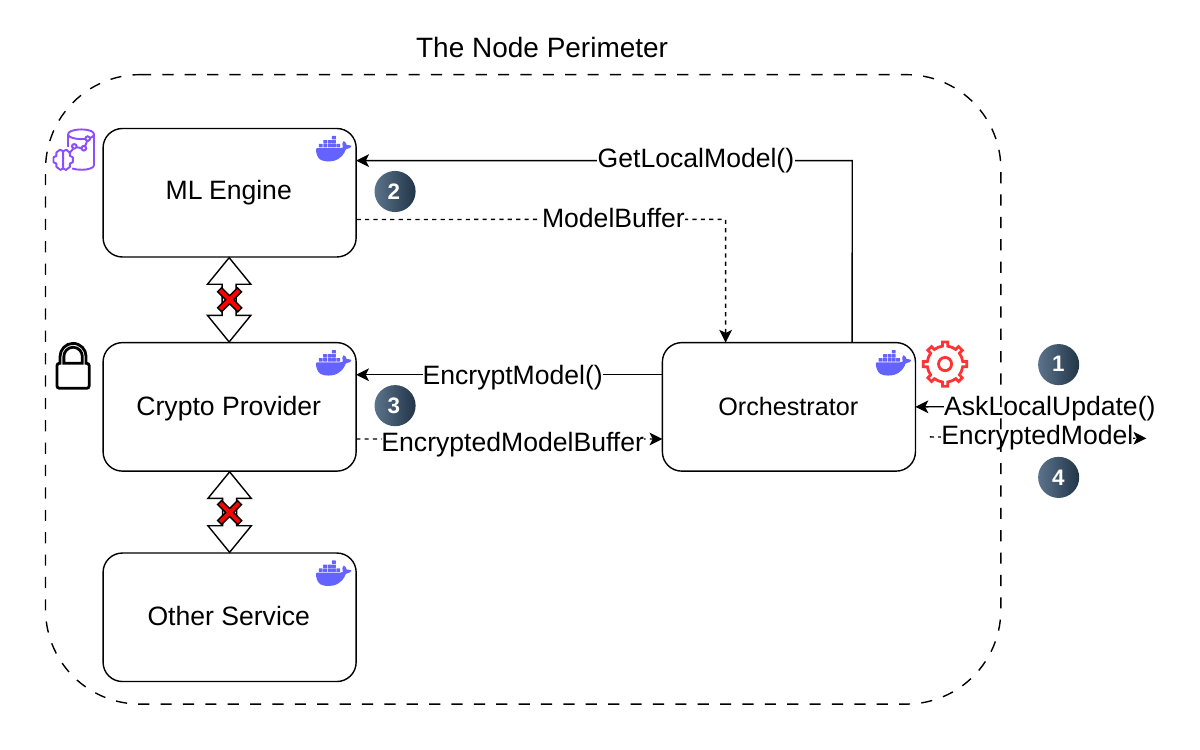}
  \caption{Micro-services architecture of a federated client node. The diagram illustrates the transactional flow of a training round: (1) the central server initiates the update request; (2) the Orchestrator performs a gRPC pull to the ML Engine to retrieve local weights; (3) the Orchestrator invokes the Crypto Provider for secure data marshaling; and (4) the encrypted payload is returned to the aggregator. Red markers indicate prohibited direct communication between sub-services, enforcing strict SoC and container-level isolation to minimize the internal attack surface. Solid lines denote gRPC control requests, whereas dashed lines represent serialized Protobuf data responses.}
  \label{fig:archi}
\end{figure}

The framework follows the principle of separation of concerns, deploying each node as a cluster of specialized, isolated containers. As illustrated in Figure \ref{fig:archi}, each node consists of three core entities: the Orchestrator, the ML Engine, and the Crypto Provider. This internal API ecosystem operates in a passive ``pull'' mode, ensuring all components remain strictly decoupled.
One might argue that this adds more complexity and communication time. Indeed, multiplying the number of interfaces and boundaries between services usually introduces latency and data serialization overhead. However, this potential bottleneck is effectively addressed by the choice of communication protocol.
To ensure high performance and scalability, the framework uses gRPC and Protocol Buffers \cite{indrasiri2020grpc}, which reduce serialization overhead compared to REST/JSON, enforce strict type safety via \texttt{.proto} service contracts, and natively handle binary payloads such as large model weights and ciphertexts.

Moreover, the microservices design enforces modularity and language-agnosticism, enabling each component to be implemented in the most suitable language, for instance Rust for the cryptography provider, and allowing users to integrate new security primitives or ML models with minimal friction.

The \textit{Orchestrator} component serves as the central intelligence and the sole gateway of the node. It acts as a proxy managing both incoming communication (from other clients or the central aggregator) and outgoing payloads. At its core, it implements a Finite State Machine (FSM) \cite{gill1962introduction} to synchronize the node's lifecycle with the global federated rounds or cryptographic protocol. The FSM is designed so that users can easily introduce new synchronization protocol features by simply modifying the state transition logic. This allows for rapid prototyping of new workflows. The \textit{ML Engine} is a passive service dedicated exclusively to tensor operations and model training. To maintain a strict separation of concerns, it remains isolated from all network and cryptographic logic. Operating strictly in ``pull'' mode, it never initiates requests; instead, it only executes tasks when prompted by the Orchestrator. This unidirectional flow ensures the machine learning workload remains entirely decoupled from the federated protocol. The \textit{Crypto Provider} handles all sensitive
cryptographic tasks, such as key generation and homomorphic operations. To ensure security, it also operates in a passive ``pull'' mode, only executing commands when prompted by the Orchestrator. While the current version implements a specific ThHE scheme, the container is designed to be easily swappable.

\subsection{Step-by-step protocol}

The operational lifecycle of MOSAIC-FL is initiated by a distributed cryptographic
setup phase in which the client nodes interact through their respective orchestrators to
jointly execute $\mathsf{Setup}(1^\lambda, N, t)$ and $\mathsf{DistKeyGen}$ procedures,
establishing the collective public key $\mathsf{pk}$ and the per-node Shamir shares
$\tilde{s}_j$ via $\mathsf{ShareReshare}$, as defined in Section~\ref{subsec:the}.
This phase involves local coordination between each node's ML and cryptographic
microservices, as detailed in Figure~\ref{fig:archi}.  To ensure model consistency
without exposing plaintext parameters to the server, participants derive the initial
global weights locally from a shared seed.

\begin{figure}[h]
    \centering
    \makebox[\linewidth][c]{%
        \includegraphics[width=1\linewidth]{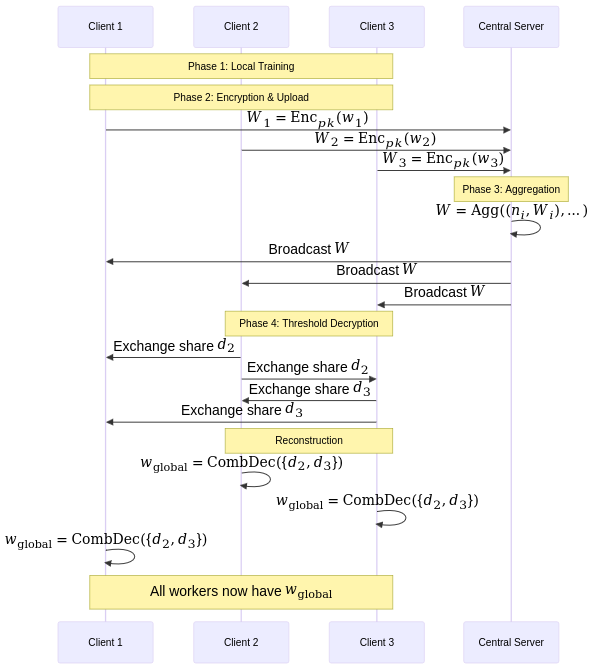}
    }
     \caption{\textbf{MOSAIC-FL Secure training round with $N=3$ total clients and $t=2$ decrypting.}}
     \label{fig:round_cycle}
\end{figure}

At the start of each training round, the server selects randomly a subset of $K$ registered
clients as active participants and notifies the remaining nodes to await the final
update. As illustrated in Figure~\ref{fig:round_cycle}, each selected client $P_k$
performs local training and its orchestrator forwards the resulting weights $w_k$ to
the cryptographic provider, which produces $\mathsf{Enc}(\mathsf{pk}, w_k)$ under
the collective public key. The central server then applies $\mathsf{Eval}$ over the received
ciphertexts and metadata to compute the $FedAvg$ aggregation function
$W = \sum_{k=1}^{K} \frac{n_k}{n}\,\mathsf{Enc}(w_k)$,
where $n_k$ denotes the local sample count of $P_k$ and $n = \sum_{k=1}^K n_k$.

Once aggregation completes, the global model $W$ is broadcast to the participant
orchestrators. Each node $P_j$ computes its partial decryption share
$d_j = \mathsf{PartDec}(W, \tilde{s}_j)$ and exchanges it with the other participants. Once at least $t$ shares are collected,
any node recovers the global model via $\mathsf{CombDec}(\{d_j\}_{j \in T})$ for
any subset $T$ with $|T| \geq t$, as described in Section~\ref{subsec:the}.
The decrypting subset $T$ may be either statically pre-configured as a fixed group
of $t$ nodes across all rounds, or dynamically designated at the beginning of each
round, where the first $t$ participants to signal readiness to their orchestrator
are selected, provided the threshold condition $|T| \geq t$ is satisfied in both
cases. Concurrently, the server opens registration for the next round,
pipelining decryption and enrollment to maximize throughput. Throughout the entire
lifecycle, secret shares and plaintext tensors remain confined within each node's
perimeter, as enforced by the container-level sandboxing described in
Section~\ref{subsec:the}. Finally, the framework supports a baseline mode that bypasses
$\mathsf{Setup}$, $\mathsf{DistKeyGen}$, $\mathsf{ShareReshare}$, and
$\mathsf{CombDec}$, serving as a cryptography-free reference for isolating
protocol overhead in the experimental evaluation of Section~\ref{sec:experiments}.

\section{\uppercase{Experimental Results}}
\label{sec:experiments}
The framework was tested for both image classification and cancer subtyping using genomic data.
\subsection{CNN-Based Federated Learning on the EMNIST Dataset}
EMNIST \cite{cohen2017emnistextensionmnisthandwritten} is a dataset composed of 800k 28$\times$28 grayscale images with 62 classes of handwritten letters and digits. The samples are allocated depending on their author and the features are rescaled. The model is a convolutional neural network (CNN) \cite{10.5555/303568.303704} made of two blocks, with a 3$\times$3 non-strided convolution and a 2$\times$2 max pooling layer each, followed by a fully-connected layer for feature reduction and a classification head, totaling 428k parameters. It is trained for 2 epochs per round, with 128-sized batches and an Adam optimizer with a $10^{-3}$ learning rate.

\subsection{Transformer for BRCA subtyping on TCGA}

Despite recent advances in FL for personalized medicine, relatively few studies have explored genomic applications, including cancer research \cite{10822609,khanna2022} or Parkinson's disease~\cite{DANEK2024100945}. Regarding secured genomics applications, fewer works have used FHE in conjunction with FL. In this context, while multi-key encryption has been proposed
 \cite{Kim2017-rk,Froelicher2021,Nguyen2025-gn}, related works have not yet explored using FHE with FL in the context of omics. A recent study~\cite{negoya2025privacypreservingfederatedlearningmethod} suggested alternatively using Differential Privacy (DP) or HE to reduce the computational overhead, nonetheless with limited security since the server can decrypt the model. This work is thus among the first efforts to use multi-key FHE applied to a genomic dataset. Here, this framework has been applied to genomics, to perform breast cancer (BRCA) subtyping using gene expression tables. BRCA samples and their corresponding PAM50 subtypes are extracted from TCGA, a database centralizing multiple cohorts~\cite{weinstein_cancer_2013}. Expression values are rescaled and class imbalance is handled with random oversampling, using scikit-learn.
Client allocation follows a target standard deviation parameter to simulate a non-IID setup. The model is a Transformer optimized for BRCA subtyping using gene expression~\cite{PAYGAMBAR2026412}. It consists of an embedding module, with a dimension reduction fully-connected layer (23$\times$20531) and a channel expansion FCN (23$\times$92), followed by four vanilla transformer blocks with four 23-sized heads, and a classification module including channel reduction and classification totaling 679k parameters. Lasso and Ridge regularizations are used with a $10^{-5}$ hyperparameter alongside a 20\% dropout. An Adam optimizer is used with a 0.0005 learning rate, and the model is trained for 50 epochs locally with a 10-epoch patience.

\subsection{Evaluation}
The experiments were conducted on a high-performance computing setup equipped with dual AMD EPYC 7302 CPUs running at 3.0GHz, providing a total of 64 cores and 128 threads, and 125GB of RAM to efficiently handle the multiple Docker containers deployed for each node to make it realistic. The system also featured an NVIDIA A100 GPU with 40GB of VRAM, enabling fast parallelized computations for deep learning tasks.
For this study, the models are implemented using TensorFlow v2.19.
All experiments using Threshold homomorphic encryption (ThHE) employed the CKKS scheme in batched mode with 128-bit security using the parameters of \href{https://github.com/tuneinsight/lattigo}{Lattigo}.
This set has a polynomial degree $N = 2^{14}$, a scaling factor $\Delta = 2^{45}$, and $PQ$ of size 438 bits with $Q$ the ciphertext modulus and $P$ the auxiliary modulus.

The performance of MOSAIC-FL was evaluated through a comparative analysis of communication overhead, model convergence, and detailed computational breakdown.

\begin{figure}[t]
    \centering
    \includegraphics [width=0.70\linewidth]{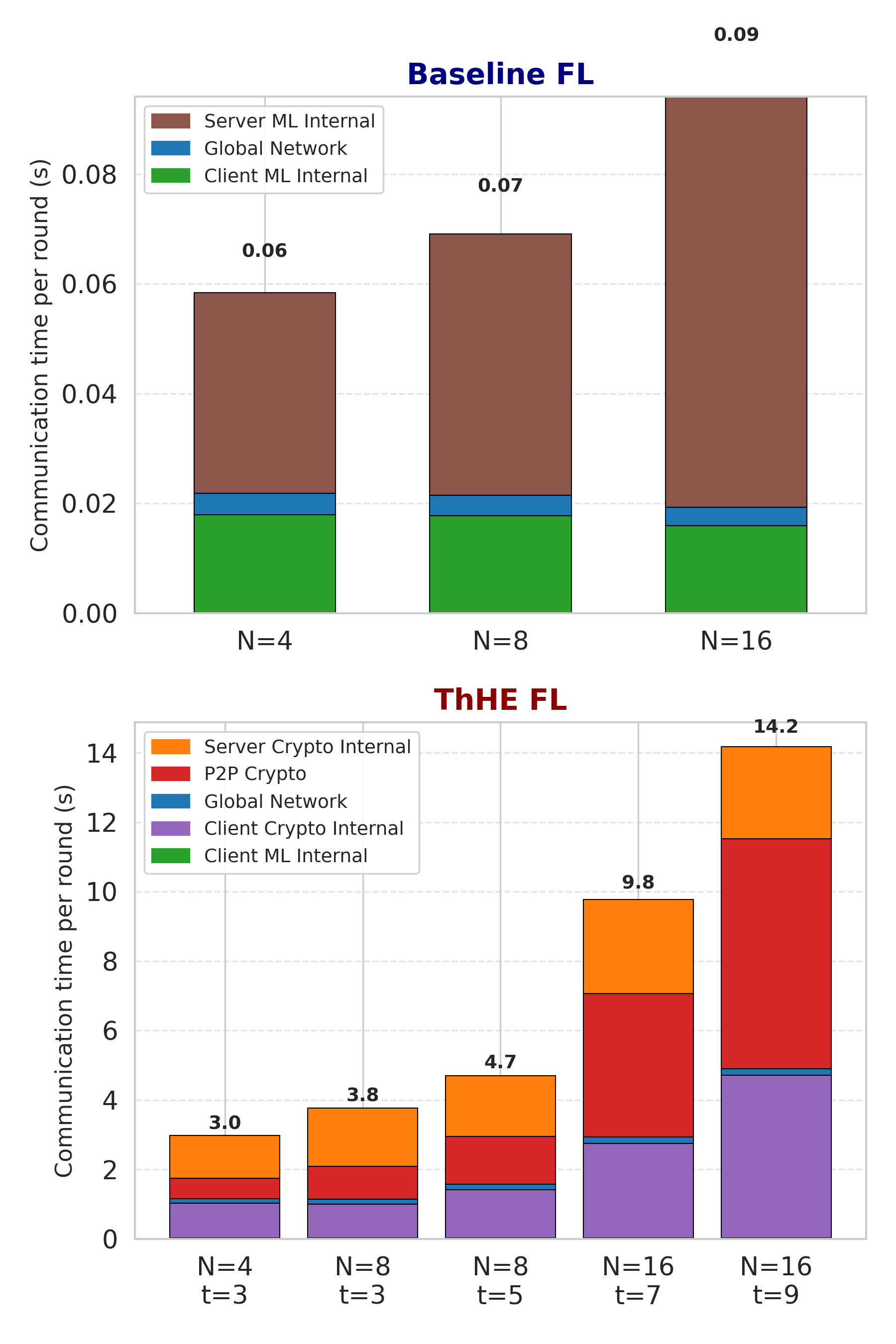} 
    \caption{\textbf{Communication overhead per round (CNN model).} Comparison between the Baseline (top) and ThHE mode (bottom) for different numbers of clients $N$ and threshold values $t$. The results highlight the impact of P2P cryptographic synchronization and secure aggregation on total latency.}
    \label{fig:comms_benchmark}
\end{figure}

The communication latency per round, illustrated in Figure~\ref{fig:comms_benchmark}, reveals a significant but manageable overhead when moving from standard FL to FL secured with ThHE-based aggregation. In the baseline FL (top) scenario, latency remains negligible (under 0.1s), representing less than 2\% of the average total transmission, idle and serialization time of 5.17s as shown in Table~\ref{tab:final_wallclock_breakdown}, with the majority of time spent on internal server ML tasks. However, the FL with ThHE (bottom) results show that as the number of clients $N$ and the threshold $t$ increase, the communication time grows significantly, reaching 14.2s for $N=16$ and $t=9$. This increase is primarily driven by the P2P Crypto component (red), which represents the collaborative cryptographic phases, notably the $\mathsf{PartDec}$ phase where clients exchange decryption shares to satisfy the threshold requirement.

\begin{figure}[t]
    \centering
\includegraphics[width=\linewidth]{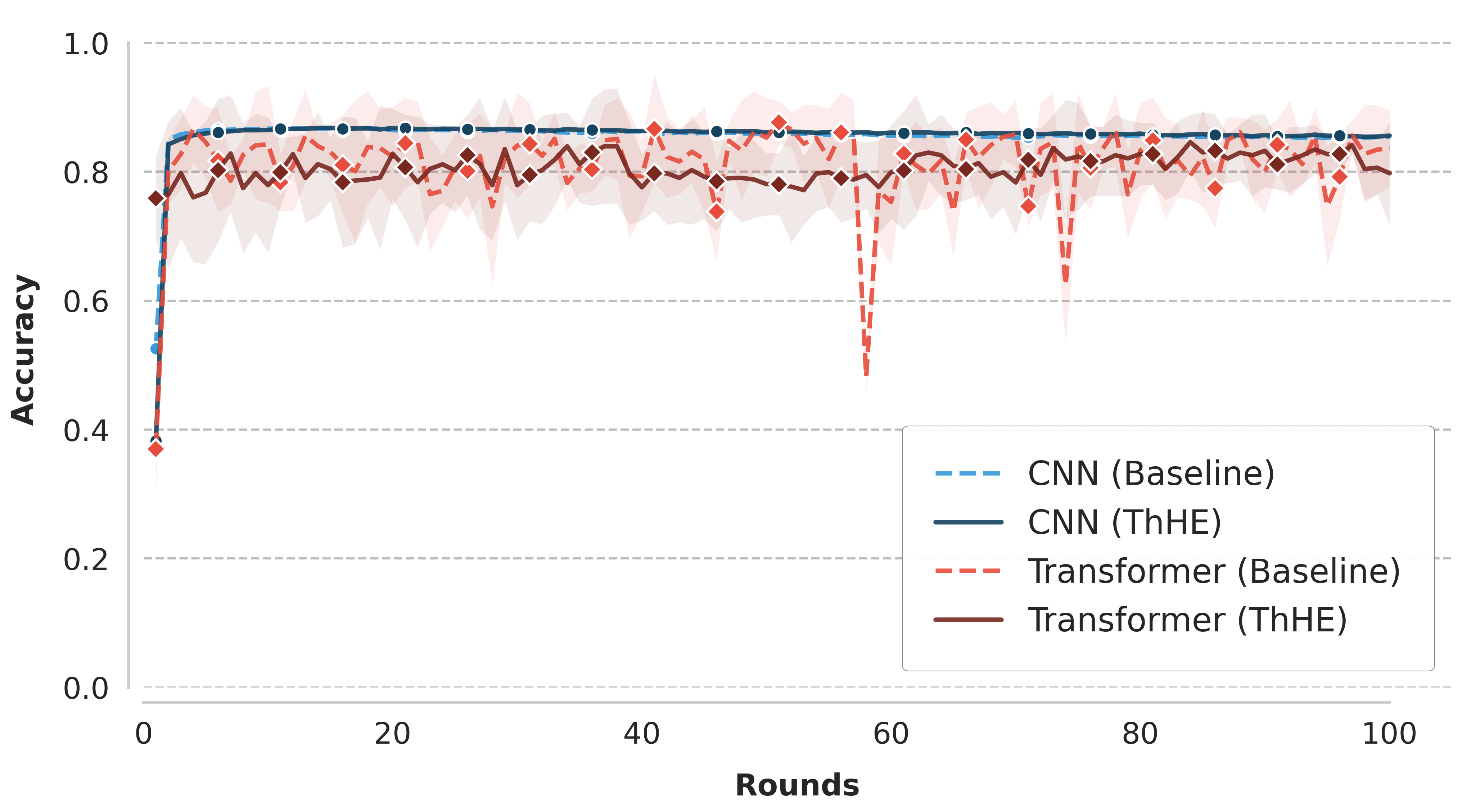}
    \caption{Accuracy evolution across 100 training rounds for CNN and Transformer use cases, with $N=4$ clients. The results demonstrate that the ThHE mode using a threshold $t=3$ consistently matches the Baseline performance. Shaded regions indicate the standard deviation among clients.}
    \label{fig:accuracy_convergence}
\end{figure}

Figure~\ref{fig:accuracy_convergence} demonstrates that the framework achieves the same level of accuracy with or without the cryptographic protocol, given an appropriate setting of the underlying security parameters.
The accuracy curves for both CNN and Transformer models in ThHE mode closely follow their non-secure counterparts. This overlap confirms that our implementation of the ThHE scheme in this framework is ``lossless,'' as the noise introduced by the CKKS homomorphic operations is properly managed and does not degrade the training performance or the final model convergence. This level of precision is primarily achieved through the configuration of the CKKS scaling factor set at $\Delta = 2^{45}$.

\begin{table*}[!t]
\centering
\caption{Detailed time breakdown per round in seconds with $N=4$ and $t=3$ (average of 100 rounds).}
\label{tab:final_wallclock_breakdown}
\footnotesize
\begin{tabular*}{\textwidth}{@{\extracolsep{\fill}} l cccccc r @{}}
\toprule
\textbf{Exp.} & \textbf{Train} & \textbf{Eval.} & \textbf{Enc.} & \textbf{Dec.} & \textbf{Agg.} & \textbf{Sync*} & \textbf{Total} \\
\midrule
CNN-Baseline & 4.40 & 0.99 & - & - & 0.002 & 5.17 & \textbf{10.56} \\
CNN-ThHE & 4.38 & 1.00 & 0.60 & 2.09 & 0.09  & 9.14 & \textbf{17.30} \\
\midrule
Trans.-Baseline & 70.96 & 0.34 & - & - & 0.004 & 87.87 & \textbf{159.17} \\
Trans.-ThHE & 69.95 & 0.24 & 0.97 & 3.26 & 0.28  & 92.35 & \textbf{167.05} \\
\bottomrule
\addlinespace[2pt]
\multicolumn{8}{l}{\scriptsize * \textit{Sync} includes network, gRPC, and idle time.}
\end{tabular*}
\end{table*}

A granular view of the different operations per round is provided in Table~\ref{tab:final_wallclock_breakdown}, where CNN-Baseline and CNN-ThHE represent the EMNIST experiments in cleartext and threshold encryption modes, respectively, while Trans.-Baseline and Trans.-ThHE denote the corresponding modes for the Transformer model. While the cryptographic overhead is evident, its impact varies significantly depending on the complexity of the machine learning task. For the CNN model, the total time per round increases from 10.56s to 17.30s when encryption is enabled. However, for the more computationally intensive Transformer model, the results demonstrate that the encryption process remains negligible in relation to the overall task. With a total time of 167.05s compared to 159.17s in cleartext, the overhead represents a minor fraction of the process, as the local training (approx. 70s) and synchronization remain the dominant factors. This highlights that for long and complex training processes, MOSAIC-FL provides robust security without compromising the overall temporal efficiency of the workflow.

\section{\uppercase{Conclusion and Future Works}}
\label{sec:conclusion}
We proposed MOSAIC-FL, a micro-service oriented secure FL framework, with a real use case for breast cancer classification. Its ability to be highly modular and to easily integrate different ML and cryptographic approaches, as well as the way it supports secure aggregation through an appropriate fault-tolerant ThHE scheme, differentiates it from existing FL frameworks. The first version was applied and validated on the well-known public EMNIST dataset as well as the more practical TCGA dataset, for a CNN and a Transformer model, respectively.
In the future, we plan to add more aggregation rules (e.g. robust or fair ones) and to extend the framework with other privacy-preserving techniques such as DP and other cryptographic modules such as VC (Verifiable Computing), in order to make it more generic and to address stronger adversaries.

\section*{\uppercase{Acknowledgements}}
This work was partially supported by the France 2030 ANR Project ANR-23-PEIA-005 (REDEEM project) under France 2030 program and by the European Union under Grant Agreement No. 101167964. Views and opinions expressed are however those of the author(s) only and do not necessarily reflect those of the European Union or the European Cybersecurity Competence Centre. Neither the European Union nor the European Cybersecurity Competence Centre can be held responsible for them.

\bibliographystyle{apalike}
{\small
\bibliography{bibliography/biblioCrypto,bibliography/biblioML,bibliography/biblioMultiKey,bibliography/gRPC}}

\end{document}